\documentclass[amsmath,amssymb,english,aps,manuscript,prl,reprint,10pt,A4,longbibliography]{revtex4-1}

\usepackage{amsmath}
\usepackage{amssymb}
\usepackage{amsfonts}
\usepackage{float}

\usepackage{graphicx}
\usepackage{subfigure}
\usepackage{color}
\usepackage{times}

\usepackage{natbib}

\newcommand{\be}{\begin{equation}}
\newcommand{\ee}{\end{equation}}
\newcommand{\ba}{\begin{eqnarray}}
\newcommand{\ea}{\end{eqnarray}}

\renewcommand{\vec}[1]{\mathbf{ #1 }}

\usepackage[T1]{fontenc}
\usepackage[latin9]{inputenc}
\usepackage{amsmath}
\usepackage{amssymb}
\usepackage{graphicx, subfigure, graphics}
\let\oldAA\AA
\renewcommand{\AA}{\text{\normalfont\oldAA}}

\begin{document}
\title{Two-dimensional epitaxial superconductor-semiconductor heterostructures: \\ A platform for topological superconducting networks}
\author{J.~Shabani$^{1}$}
\author{M. Kjaergaard$^{2}$}
\author{H.~J. Suominen$^{2}$}
\author{Younghyun Kim$^{3}$}
\author{F. Nichele$^{2}$}
\author{K. Pakrouski$^{4}$}
\author{T. Stankevic $^{2}$}
\author{R.~M.~Lutchyn$^{6}$}
\author{ P. Krogstrup$^{2}$}
\author{ R. Feidenhans'l$^{2}$}
\author{S. Kraemer$^{5}$}
\author{C. Nayak$^{3,6}$}
\author{M. Troyer$^{4}$}
\author{C. M. Marcus$^{2}$}
\author{C. J. Palmstr\o m$^{1,5,7}$}
\affiliation{$^{1}$California NanoSystems Institute, University of California, Santa Barbara, CA 93106, USA
\\
$^{2}$Center for Quantum Devices and Station Q Copenhagen, Niels Bohr Institute, University of Copenhagen, 2100 Copenhagen, Denmark
\\
$^{3}$Department of Physics, University of California, Santa Barbara, CA 93106, USA
\\
$^{4}$ Theoretical Physics and Station Q Zurich, ETH Zurich, 8093 Zurich, Switzerland
\\
$^{5}$Materials Research Laboratories, University of California, Santa Barbara, CA 93106, USA
\\
$^{6}$Microsoft Research, Station Q, University of California, Santa Barbara, CA 93106, USA
\\
$^{7}$Department of Electrical Engineering, University of California, Santa Barbara, CA 93106, USA
}
\date{\today}
\maketitle

\textbf {Progress in the emergent field of topological superconductivity relies on synthesis of new material combinations, combining superconductivity, low density, and spin-orbit coupling (SOC). For example, theory \cite{Sau, Alicea, 1DwiresLutchyn, 1DwiresOreg} indicates that the interface between a one-dimensional (1D) semiconductor (Sm) with strong SOC and a superconductor (S) hosts Majorana modes with nontrivial topological properties \cite{TQCreview, Kane_review, Alicea'12, Beenakker_review}. Recently, epitaxial growth of Al on InAs nanowires was shown to yield a high quality S-Sm system with uniformly transparent interfaces \cite{Peter2015} and a hard induced gap, indicted by strongly suppressed subgap tunneling conductance \cite{Willy2015}. Here we report the realization of a two-dimensional (2D) InAs/InGaAs heterostructure with epitaxial Al, yielding a planar S-Sm system with structural and transport characteristics as good as the epitaxial wires. The realization of 2D epitaxial S-Sm systems represent a significant advance over wires, allowing extended networks via top-down processing. Among numerous potential applications, this new material system can serve as a platform for complex networks of topological superconductors with gate-controlled Majorana zero modes \cite{Sau, Alicea, 1DwiresLutchyn, 1DwiresOreg}. We demonstrate gateable Josephson junctions and a highly transparent 2D S-Sm interface based on the product of excess current and normal state resistance.}

The recent focus on topological states in solid state systems has revealed new directions in condensed matter physics with potential applications in topological quantum information \cite{ReadGreen, Kitaev:2001}. In an exciting development, it was realized one could readily engineering an effective one-dimensional (1D) spinless superconductor using the proximity effect from conventional superconductors (Al, Nb) in nanowires with strong SOC (InAs, InSb), and that Majorana zero modes would naturally emerge at the ends of the wire \cite{1DwiresLutchyn, 1DwiresOreg, Sau, Alicea}. First experiments on nanowires grown by chemical vapor deposition (CVD) revealed striking evidence of Majorana zero modes states~\cite{Mourik2012, Rokhinson2012, Das2012, Deng2012, Fink2012, Churchill2013}. In order to eventually move beyond demonstrations of braiding ~\cite{AliceaNatPhy11, ClarkePRB11, HalperinPRB12}, to larger-scale Majorana networks~\cite{AliceaBraiding} it is likely that a top-down patterning approach will be needed. Molecular beam epitaxy (MBE) growth of large-area 2D S-Sm systems can form the basis for such an approach, but to date have not been available.

\begin{figure*}[htp]
\centering
\includegraphics[scale=0.45]{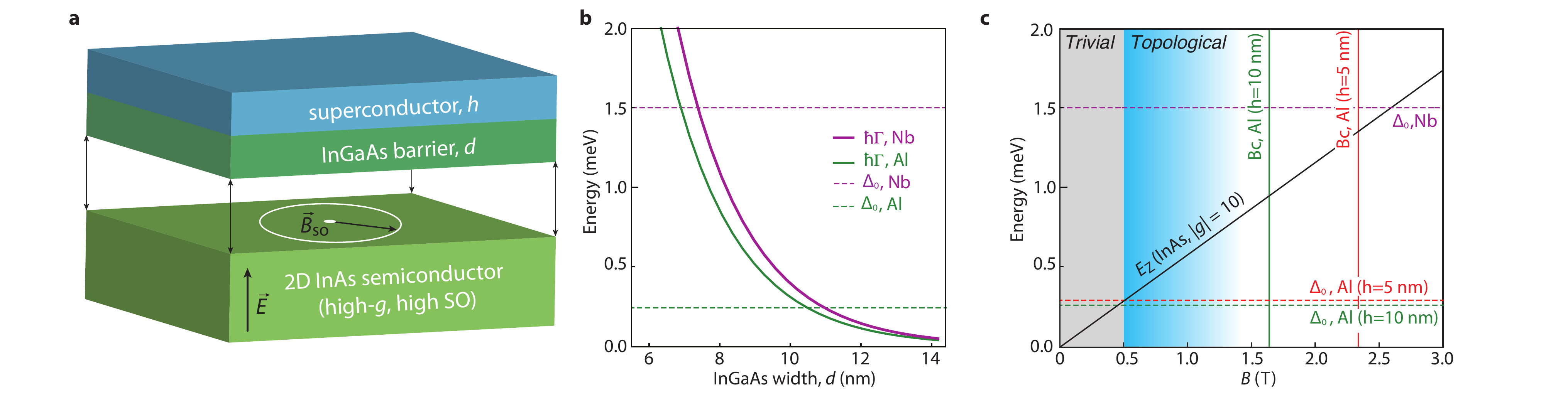}
\caption{(Color online) (a) Proposed structure for two-dimensional superconductor-semiconductor interface. The sketch is exploded in the barrier/InAs interface to highlight the spin--orbit field in the plane of the InAs. (b) Tunneling rate of electrons, $\hbar\Gamma$, for Al and Nb as a function of top barrier thickness, $d$, see text. (c) Plot of Zeeman energy, $E_{Z}$, as a function of $B$ for InAs ($|g| = 10$). Critical fields and superconducting gaps of Al for two thicknesses of 5 and 10 nm are also shown.}
\end{figure*}

Narrow bandgap semiconductors such as InAs and InSb are natural choices for the Sm component due to large $g$ factors and strong SOC, which are important for the stability of an emergent topological phase in S-Sm heterostructures, with the topological gap proportional to the SOC strength~\cite{ZeroBiasAnomaly31}. There are, however, significant challenges in growing high quality quantum wells in these systems. The lack of insulating lattice-matched substrates and difficulty in device fabrication, compared to well-developed GaAs material system, has restricted their use in mesoscopic devices. Nevertheless, it has long been known \cite{Clark1980} that surface level pinning in InAs could allow for fabrication of transparent contact to superconductors and high quality S-Sm-S devices have been reported using in-situ ion milling of the native oxide \cite{Richter_1999, Kroemer_1994}. In this work we adopt a different approach by growing epitaxial layers of Al on 2D InAs/InGaAs quantum wells. These systems represent the ideal scenario in achieving a flat, abrupt and impurity-free interface. We show that our material system, Al-InAs, satisfies all the requirements necessary to reach the topological superconducting regime.

The recipe for creating a hybrid system that supports topological superconductivity requires a balance between proximity and segregation of constituent materials \cite{TooMuchofaGoodThing}. The interface must allow electrons to inherit superconducting correlations from the s-wave superconductor while retaining large SOC and large $g$-factor from the semiconductor. This balance depends on how the electron wave function resides in both materials. Theory~\cite{ZeroBiasAnomaly61} suggests that the average time spent by a quasiparticle in the Sm region is determined by the hybridization with the metallic states in the S region, $1/\Gamma$ (i.e., escape time from a quantum well to the normal metal) whereas the average time spent in the S region is given by the Heisenberg uncertainty time $\hbar/\Delta$ with $\Delta$ being the quasiparticle gap. An optimal balance is achieved when $\hbar\Gamma \sim \Delta$, i.e., when a quasiparticle spends roughly equal time in the S and Sm regions. Thus, in order to realize robust topological superconductivity, it is not only important to achieve highly transparent and disorder-free contacts between the active electrons in the Sm and the S, but also necessary to tune the tunneling between Sm and S regions with a barrier. This could be achieved, for instance, by inserting a potential barrier (e.g., a layer of InGaAs) between Sm and S. A calculation of $\hbar\Gamma$ versus barrier thickness $d$ is shown in Fig.~1b for the case of Al and Nb as S and InAs as Sm materials. Because of the different bulk superconducting gaps, $\Delta_{0}$, and different Fermi energies, optimal barrier thicknesses differs in the two cases.

A quantum phase transition from trivial to topological superconducting state can be driven by an external magnetic field, $B$~\cite{Alicea}. This requires a superconductor that can tolerate magnetic fields exceeding $\Delta/(g\mu_B)$, where $\mu_{B}$ is the Bohr magneton and $g$ is the g factor in the semiconductor \cite{Sau, Alicea'12}. Bulk aluminum has a critical field $B_c$ of the order of $50~\rm{mT}$, too low to drive the system in the topological regime, even with $g \sim 10$ or larger in the semiconductor. However, few-nm thick Al film can sustain in-plane fields in excess of 2 T, readily exceeding $\Delta/(g \mu_{B})$  \cite{TedrowMes82}. Figure 1c plots two energies, $\Delta$ and $E_{Z} = g\mu_B B$ for Al. The in-plane critical fields for Al depends on the thickness of Al films. We note that material structure such as Fig. 1a, where the confinement potential is highly asymmetric, can enhance SOC over the bulk values, leading to a larger quasiparticle gap. 

\section{High mobility, high spin-orbit, near surface two-dimensional electron system}

\begin{figure*}[htp]
\centering
\includegraphics[scale=0.75]{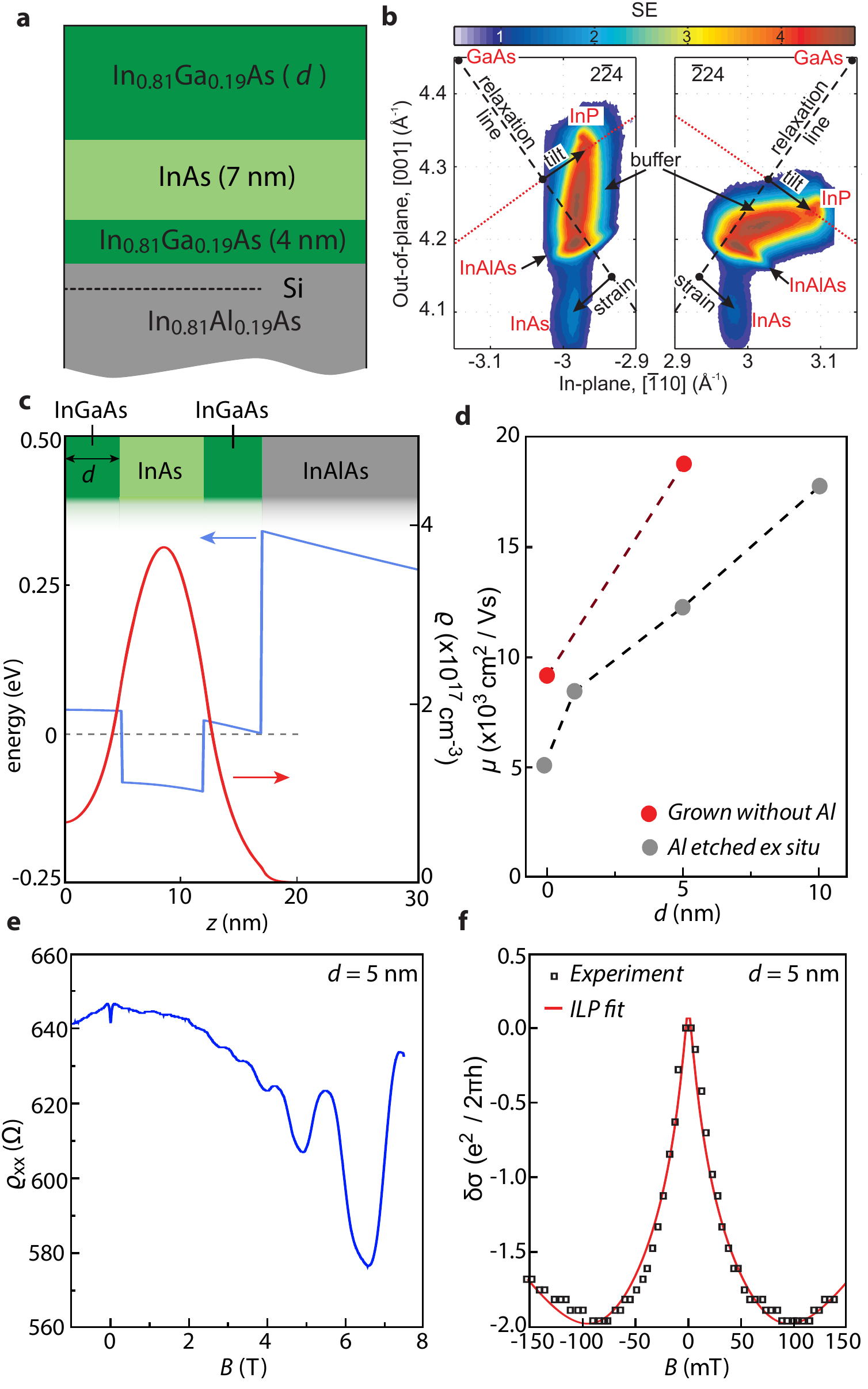}
\caption{(Color online) (a) Layer structure of our InAs quantum well near surface. (b) Reciprocal space maps of the two opposite Bragg peaks for the semiconductor: 2-24 and -224. The dashed line is the relaxation line and the red dotted line is the Debye-Scherrer ring. (c) Charge distribution calculated for $d = 5$ nm. (d) Electron mobilities as a function of top barrier thickness, $d$. Red (Black) symbols show structures that are grown without Al (with Al and then etched). (e) Magneto-transport measurements of $d$ = 5 nm at $n = 1 \times 10^{12}$ cm$^{-2}$. (f) Weak antilocalization signal, $\delta \sigma = \sigma (B) - \sigma (B = 0)$, near zero magnetic field. The red is the ILP fit to the data, see text. }
\end{figure*}

We first present the structural and electronic properties of our near surface InAs quantum wells. Figure 2a shows the schematics of the material stack \cite{Richter2000, ShabaniAPL2014, ShabaniMIT}. The structures are grown on a semi-insulating InP (001) substrate with In$_{x}$Al$_{1-x}$As buffer where the indium content is step graded from $x =$ 0.52 to 0.81. The quantum well consists of In$_{0.81}$Ga$_{0.19}$As and InAs layers. X-ray diffraction analysis shows that the upper functional layers of samples are typically tilted with respect to the InP substrate from 0.15(2)$^{\circ}$ to 0.8(2)$^{\circ}$. It also exhibits isotropic mosaicity of the InAlAs layer in the range of 0.4$^{\circ}$ at FWHM, originating from the cross hatched pattern due to strain relaxation by dislocations.

Reciprocal space maps (RSM) of the (2-24) and (-224) Bragg peaks of the semiconductor are shown in Fig. 2b, The maps are aligned with [-110] and [001] directions of InAlAs on the axes. A smooth transition is evident from InP to InAlAs through the graded buffer. However, there is a notable asymmetry of InP peak position in the two RSMs. The InP peaks are shifted from the relaxation line clockwise along the Debye-Scherrer ring (red dotted line in Fig.~2b), which corresponds to the crystal tilt between the layers above the buffer and the underlying substrate. The two InGaAs layers and InAlAs have very similar lattice constants; therefore we cannot distinguish between them and they all contribute to the peak labeled as InAlAs. Strain and composition of In$_{x}$Al$_{1-x}$As were calculated from the peak positions assuming Vegard's law and using bulk lattice parameters and elastic constants of InAs and AlAs. The InAs layer is seen as the weakest peak at lowest out-of-plane Q values. It is fully strained with respect to the underlying InGaAs and InAlAs, which corresponds to compressive in-plane strain of $\epsilon_{xx}$=1.6(2)\% (with a corresponding out-of plane strain of $\epsilon_{zz}$=1.1(2)\%, consistent with reported values for Poisson's ratio).

The surface InAs quantum wells have relatively low electron mobilities (under 10,000 cm$^2$/V\,s), mostly due to the direct contact of electrons to scattering impurities at the S-Sm interface. A top barrier (thickness $d$) can improve the situation by separating the quantum well from the Al interface. InAlAs barriers could be used, but would likely result in a too abrupt wave function confinment, not allowing sufficient overlap with the S region. InGaAs is a more suitable choice because the smaller electron mass increase the length over which the wave function decays in the barrier region.  Electron density distribution with an InGaAs barrier, calculated using a self-consistent Poisson-Schrodinger solver, is shown in Fig.~2c. As $d$ is increased the charge distribution is moved away from the surface resulting in a mobility increase. Figure 2d shows the summary of the sample mobilities as a function of the InGaAs top barrier, $d$. Two sets of data are shown on wafers without in-situ growth of Al (S) and with Al but removed after growth using a selective wet etch. We find that, even for the same $d$, the mobility of electrons is higher when Al is not chemically etched. This indicates that the surface treatments are crucial and special care must be taken to ensure no mobility degradation. A possible way to avoid this chemical reaction is the full oxidation of the Al film.

\begin{figure*}[htp]
\centering
\includegraphics[scale=0.75]{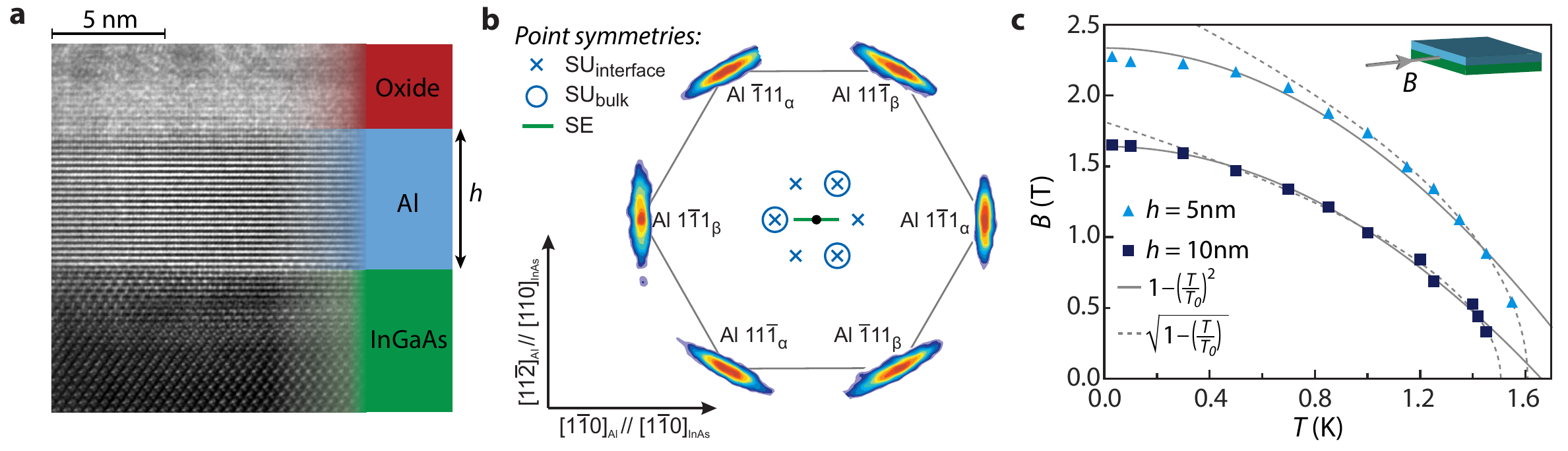}
\caption{(Color online) (a) High-resolution transmission electron microscope image showing that the Al
forms a sharp and uniform interface to the InGaAs layer. (b) Reciprocal space maps for the [111] out-of-plane orientation shows six non-specular (111) peaks indicating that two grain orientations are present. For details on the degenerate interfacial grain orientations, see Ref. \cite{Peter2015}. (c) Critical temperature and magnetic field measurements for a 5 nm (triangles) and 10 nm (squares) Al films on In$_{0.81}$Ga$_{0.19}$As/InAs. The solid and dashed curves are the scaling fits.}
\end{figure*}

In a perpendicular magnetic field, the system exhibits Shubnikov-de Haas (SdH) oscillations in the $d$ = 5 nm wafer with an onset of oscillation about 2 T as shown in Fig. 2e. The weak antilocalization in this wafer is analyzed using the theory developed by Iordanski, Lyanda-Geller, and Pikus (ILP) for two-dimensional electron systems (2DESs) \cite{ILP, Knap96}. The theory is valid when either Rashba or linear Dresselhaus SOC is dominating the other. To reduce the number of free fitting parameters we fixed the value of cubic Dresselhaus SOC, $\gamma$ as the bulk value of InAs 26.9 eV $\AA^3$ calculated from the $\vec{k} \cdot \vec{p}$ theory \cite{Knap96, SemiBook}. The resulting linear Dresselhaus SOC for our 2DES can be estimated using $\alpha_D = \gamma (\left<k_{z}^{2}\right>-\frac{1}{4}k_{F}^{2})$, where $\left<k_{z}^{2}\right>$ is the average squared wavevector in the growth direction $z$, to be $\alpha_{D} \sim$ 50 meV $\AA$ \cite{Knap96}. The remaining fit parameters are phase-coherence length, $l_{\phi}$, and linear spin-orbit coupling $\alpha$. Fitting $\delta \sigma (B)$ over the range $|B|<150$ mT at T = 2 K yields parameters $l_{\phi}$ = 350 nm and $\alpha$ = 280  meV $\AA$. The fact that $\alpha > \alpha_D$ indicates that Rashba SOC is the dominant contribution. This value of $\alpha$ corresponds to a spin-orbit length $l_{so} = 45$ nm, and $l_{\phi}/l_{so} \sim 8$.


\section{Epitaxial growth of Superconductor on 2DES}

Figure 3a shows a high-resolution transmission electron microscope (TEM) image of epitaxial Al on In$_{0.81}$Ga$_{0.19}$As (001), with atomic planes of both crystals clearly visible.  X-ray diffraction (XRD) studies only show Al (111) out-of-plane orientations. The azimuthal orientation of Al(111) with respect to the underlying semiconductor is determined by making a full sample rotation while measuring asymmetric Al $\{111\}$  peaks, as illustrated in Fig. 3b. Six equally spaced Al $\{111\}$ peaks with equal intensity are found. Since the point symmetry of the InGaAs surface is two-fold (indicated with a line) and the corresponding symmetry of the Al bulk is three-fold, rotating the Al implies two degenerate interfacial configurations. The same degeneracy would appear if the structure was solely determined by single plane interfacial bicrystal symmetries, where the Al (111) interface is 6-fold (indicated with 6 `$x$' symbols around the center point). This means that the interface consist of only one type of interfacial bonding, indicating a strong two-fold degenerate minimum with only the lowest number of grain types possible. By measuring the distances between the opposite peaks we can conclude that Al is relaxed (unstrained) within measurement uncertainty ($\pm$ 0.1\%).

The growth mechanisms of the Al film can be described in the thin-film limit (of film thickness $h$), where the size-dependent part of the chemical potential of a S grain with in-plane radius of curvature $R$, is given by \cite{Peter2015}:

\begin{equation}
\delta \mu_{S}^{R} \propto \frac{\gamma_{S}}{h}+\frac{\gamma_{Sm||S}}{h}+\frac{\gamma_{\bar{S}||S}}{R}+\frac{G \epsilon^{2}}{(1-\nu)}
\end{equation}

 The four terms in Eq.~(1) account for the free energy excesses of the surface, the S-Sm interface, the grain boundaries and the strain energy, respectively, where $\gamma_{S}$, $\gamma_{Sm||S}$ and $\gamma_{\bar{S}||S}$ are the corresponding excess free energies due to the chemical bonding of the interfaces, $G$ is the shear modulus, $\epsilon$ the strain, and $\nu$ Poisson's ratio. In the thin film limit, $h \ll R$, mechanisms determining in-plane and out-of-plane and crystal orientations can be separated. The strongest thermodynamic driving force is the surface free energy minimization ($\gamma_{S}>\gamma_{Sm||S}$), which determines the out-of-plane orientation. This is typically (111) for FCC materials, including Al. The in-plane orientations are secondarily determined by the last three terms and involve more considerations \cite{Peter2015}. At the initial stage of the growth, when $h$ is sufficiently small, the chemical bonding at the S-Sm interface (i.e. second term in Eq.~(1)) dominates and dictates the in-plane orientation. In the x-ray and TEM measurements, we observe two rotational grains but the same interfacial structure across the wafer. As the Al thickness approaches the critical value, given by, $h_{c}=\gamma_{Sm||S}(1-\nu)/(G\epsilon^{2})$, where the strain energy exceed the difference in chemical bonding energy between the strained (domain matched) and the relaxed Al, the film will start to relax. We conclude that $h_{c} < 5$ nm as 5 nm thick Al films are determined to be relaxed.

If the indium composition in In$_{x}$Ga$_{1-x}$As, $x$, is varied, the lattice constant of the semiconductor changes and the strain energy density $\frac{G\epsilon^{2}}{(1-\nu)}$ for a given domain match will change. Here $\epsilon^{2} = \epsilon_{1}^{2}+\epsilon_{2}^2+2\nu\epsilon_{1}\epsilon_{2}$ , where $\epsilon_{1}$ and $\epsilon_{2}$ are the strain in the two in-plane directions. If the strain energy for a given domain match is too high, $h_{c}$ might be smaller than the thickness of the initial nucleus, and the film will either find a different lower symmetry match or appear more disordered on a macroscopic scale with many different grain orientations. Strain energy calculation as a function of indium content, $x$, exhibits a minimum energy near $x \sim 0.8$. This suggest that growth of Al on In$_{0.8}$Ga$_{0.2}$As results in a smoother interface consistent with TEM images near this composition.

How grain boundaries affect the electronic properties of the S-Sm interface is poorly understood. Figure~3c shows a comparison of the critical magnetic field as a function of  temperature for Al on InGaAs barriers. Critical magnetic fields at base temperature, $T$ = 30 mK, are found to be $B_{c}(0) $= 2.3 (1.6) T for 5 (10) nm films. At elevated temperatures, the critical field data are reasonably well fit by the Bardeen-Cooper-Schrieffer (BCS) form $B_{c}(T)=B_{c}(0)[1-(T/T_{c})^{2}]$ \cite{Tinkham:2004un} taking $T_{c}$ and $B_{c}$ as fitting parameters. At the low-temperature end of the scale, the 5 nm Al is better described by Chandrasekhar-Clogston theory~\cite{Chandrasekhar1962, Clogston} where the upper limit critical field is expected to reach $B_{c} = \Delta_{0}/\sqrt{2} \mu_{B} \sim 2.4$ T. Close to $T_{c}$, data for both thicknesses fit the form $B_{c}(T)=B_{c}(0) \sqrt{1-T/T_{c}}$ \cite{Wojcik2015}.

\section{Gateable supercurrent}

Having shown that the Al-InAs platform satisfies several basic requirements for topological superconductivity, we next demonstrate proximity effect and gate control in an S-Sm-S geometry. This geometry provides a probe of S-Sm interface transparency.  High interface transparency, corresponds to a high probability of Andreev reflection at the S-Sm interface, is reflected in the supercurrent through the S-Sm-S structure.

Figure 4a shows a scanning electron micrograph of an S-Sm-S device with barrier $d$ = 10 nm and Al thickness $h$ = 10 nm. Selective etching has been used to remove a thin strip of aluminum, followed by deposition of 40 nm of aluminum oxide by atomic layer deposition (ALD) and a metallic top gate. The junction is 3$~\mu$m wide and has a 200~nm separation between the superconducting electrodes. The I-V characteristic of the junction is measured at 30 mK in Fig.~4b. The voltage drop across the junction is zero (the supercurrent) up to a critical value of driving current denoted the critical current, $I_{c} = 1.4 ~\mu$A (see Fig.~4c). As the gate is used to deplete the 2DES, the critical current remains nearly unchanged down to $V_g < -2~$V. At more negative gate voltages the critical current is reduced, roughly inversely proportional to the above-gap resistance. The above-gap resistance is approximately equal to the normal state resistance, $R_{n}$. For gate voltages in the range  $-3~V<V_g <-2~V$ the gate voltage decreases while the product $I_{c}R_{n}$ remains roughly constant, and in fact slightly increases (Fig.~4d). For  $V_g <-3\,V$ the $I_{c}R_{n}$ decreases rapidly as the critical current vanishes with the junction becoming insulating \cite{ShabaniMIT}. 

Transport measurements on Hall bars with the Al removed at $V_g = 0~$V yield a mean free path of $l_{e} \sim$ 230$~$ nm, indicating that the junction is neither clearly ballistic nor diffusive. Note that in the present geometry, the Sm extends under the S regions. The interface between Sm and S is highly transparent due to the large area of contact and \emph{in situ} aluminum growth. The Andreev process that carries the supercurrent across the Sm region is characterized by the induced gap $\Delta_{\rm ind}$ in the Sm below the S rather than the bulk Al gap, $\Delta_0$. To characterize an S-Sm-S junction in the short limit the product of the critical current and the normal state resistance, which is related to the gap $I_{c}R_{n} = a \Delta_{0}/e$, is often used. Here, $a$ is a parameter of order unity and is model dependent \cite{Cheng12}. We find $I_{c}R_{n} = 135~\mu $V at $V_g = 0~$V in our device which is close to the bulk gap of the Al thin film at this thickness, $\Delta_{0}/e \sim 200 ~\mu$V. To our knowledge, previous studies of ex-situ fabricated junctions on 2DES have reported $I_cR_n$ products typically an order of magnitude smaller than $\Delta_{0}$ \cite{Nitta92,Takayanagi95,Mur96,Heida98}.

\begin{figure*}[htp]
\centering
\includegraphics[scale=0.75]{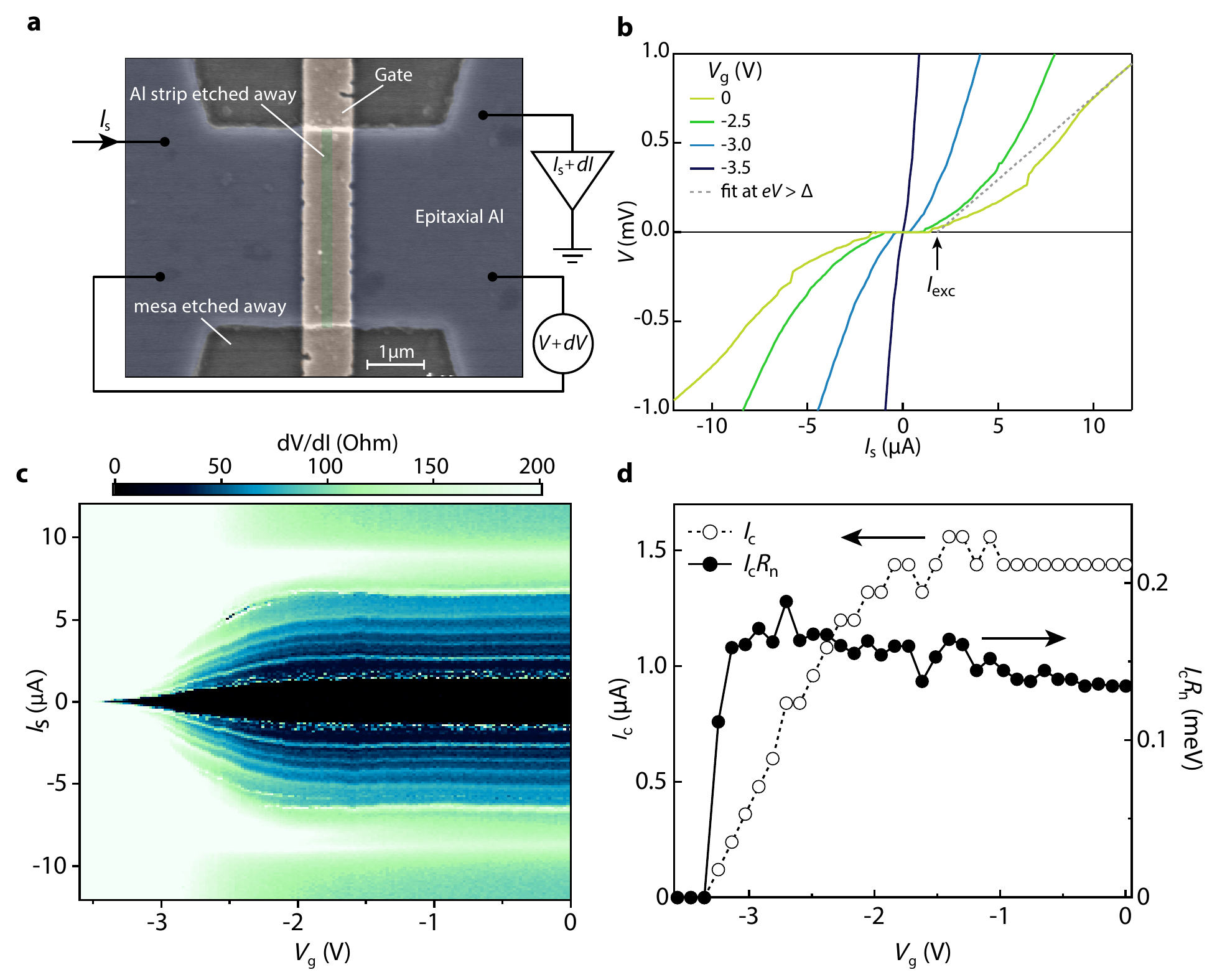}
\caption{(Color online) (a) A scanning electron microscope (SEM) image of a gated S-Sm-S junction fabricated on a wafer with d = 10 nm top barrier (b) $V-I$ characteristics of the junction as a function of top gate. An example of a linear fit to determine excess current is shown in dashed line for $V_g=0~$V. (c) $dV/dI$ vs dc source current, $I_{s}$ measured as a function of top gate voltage. (d) The gate dependence of the critical current, $I_{c}$, and the product of the critical current and the normal state resistance, $I_{c}R_{n}$. }
\end{figure*}

Each Andreev reflected electron contributes $2e$ to the current through the junction, leading to an excess current relative to a normal metal junction. This excess current is thus an indirect measurement of the quality of the S-Sm interface, and is found by extrapolating a linear fit at $eV > \Delta_0$ to $V = 0~$V  \cite{Blonder:1982bc}. An example of the procedure is shown as a dashed gray line for $V_g = 0~$V in Figure 4b, and we find $I_{exc} = 1.99 ~\mu$A. We assume a diffusive junction \cite{Giazotto:2004a}, for which the excess current through a perfect S-Sm interface is related to the gap via $I_{exc}R_{n} = (\pi^2/4 -1)\Delta_{0}/e$~\cite{Artemenko:1979kb,Athanassios:1997a}. Using $\Delta_{0}$ (instead of the unknown $\Delta_{\rm ind}$) of our aluminum film, an upper bound for the induced gap in the semiconductor\cite{Volkov:1995ip,Aminov:1996uj,Chrestin:1997dh}, we obtain $I_{exc}R_{n} = 330~\mu $V. Experimentally we measure $I_{exc}R_{n} = 191~\mu$V in close agreement with the theoretical predictions for an ideal interface.

\section{conclusion}

In this work, we demonstrated that epitaxial Al-InAs two-dimensional systems are a viable platform to study topological superconductivity. We used InGaAs top barriers to achieve high electron mobilities and facilitate the growth of ultra-thin film Al. The electronic and material properties of both 2DES and Al are characterized in 2D and in S-Sm-S junctions. We observe an exceptional quality of S-Sm junctions compared to earlier experiments that attest to the quality of the S-Sm interface. Fabrication of complex architectures offers endless possibilities for exploring new directions. Our findings are expected to spark interest in large-scale device applications in mesoscopic and topological superconductivity.

\section{Methods}

The samples were grown on a semi-insulating InP (100) substrate, using a modified VG-V80H molecular beam epitaxy system. After oxide desorption under an As$_{4}$ overpressure at 520 $^{\circ}$C, the substrate temperature is lowered to 480 $^{\circ}$C where we grow a superlattice of InGaAs/InAlAs lattice matched to InP. The substrate temperature is further lowered to 320 $^{\circ}$C for growth of the In$_{x}$Al$_{1-x}$As buffer layer. This lower temperature buffer growth is used to reduce and minimize the influence of dislocations forming due to the lattice mismatch of the active region to the InP substrate. The indium content in In$_{x}$Al$_{1-x}$As is step graded from $x =$ 0.52 to 0.81. We keep the indium content at $x =$ 0.81 while increasing the substrate temperature to $T_{sub} \sim$ 400 $^{\circ}$C with increasing As$_{4}$ flux to observe a $2 \times 4$ surface reconstruction. The quantum well consists of a 7 nm InAs grown on a 5 nm In$_{0.81}$Ga$_{0.25}$As. The structures are Si delta doped, $N_{D} = 1 \times 10^{12}$ cm$^{-2}$, 7 nm below the InGaAs layer. Following InAs growth, the substrate temperature is kept at 200 $^{\circ}$C until all the sources are ramped down to their idle temperatures. The substrate is then cooled by turning off all power sources that can act as heat sources (power supply for substrate holder, ion gauges, light sources). This process typically takes more than 6 h in our chamber. Three dimensional reciprocal space maps were measured using the 12.4 keV x-ray photon energy at I811 beamline at MAX II synchrotron radiation facility in Lund, Sweden.

\section{Acknowledgements}

Research supported by Microsoft Research, the Danish National Research Foundation, the Swiss National Science Foundation through QSIT, and the US NSF through the National Nanotechnology Infrastructure Network. F.N. acknowledges support of the EC through the Marie Curie Fellowship. R.L. acknowledges the hospitality of the Aspen Center for Physics supported by NSF grant PHY-1066293, where part of this work was done. We acknowledge contributions of the beamline staff at I811 MAX II synchrotron.

\bibliography{topological_wires}


\clearpage

\widetext
\begin{center}
\textbf{\large Supplemental Material for "Two-dimensional epitaxial superconductor-semiconductor heterostructures: \\ A platform for topological superconducting networks"}
\end{center}

\setcounter{equation}{0}
\setcounter{figure}{0}
\setcounter{table}{0}
\setcounter{page}{1}
\makeatletter
\renewcommand{\theequation}{S\arabic{equation}}
\renewcommand{\thefigure}{S\arabic{figure}}

\section{Semiconducting material properties}

The density of the two-dimensional electron system (2DES) can be controlled by depositing a gate on aluminum oxides. The oxides are typically 40 nm thick grown using atomic layer deposition (ALD) system. The electron densities measured in Hall bars as a function of top gate show a change of density $10^{16}~\text{m}^{-2}$ upon application of $1~V$. The electron mass has been measured in similar structures to be $m_{e} = 0.03~m_{0}$ using the temperature dependence of Shubnikov-de Haas oscillations at filling factor $v $ = 12. A $|g|$ factor of 10 is deduced from magnetomeasurement in tilted magnetic field.

\section{Growth of Al thin films}

Following our discussion in the text, the indium composition in In$_{x}$Ga$_{(1-x)}$As, $x$, can be varied to detemine the domain matching with lowest strain energy density (responsible for the fourth term in Eq.~(1) in the main text). Using vegards law, varying $x$ results in a linear change in the lattice constant of the semiconductor and the
strain energy density $\frac{G\epsilon^{2}}{(1-\nu)}$ will change accordingly for a given domain match \cite{Peter2015}. Figure S1(a) shows a normal view of the interfacial domain match of strained and relaxed Al-In$_{0.8}$Ga$_{0.2}$As. It is important to note that the relaxed situation is only to show the mismatch of the two phase in relaxed state, the interface is at relaxed bulk conditions, most likely associated with a dislocation array at the interface with respect to a given domain matching with the lowest chemical bonding energy (second term in Eq.~(1)).

\begin{figure}[H]
\centering
\includegraphics[scale=1]{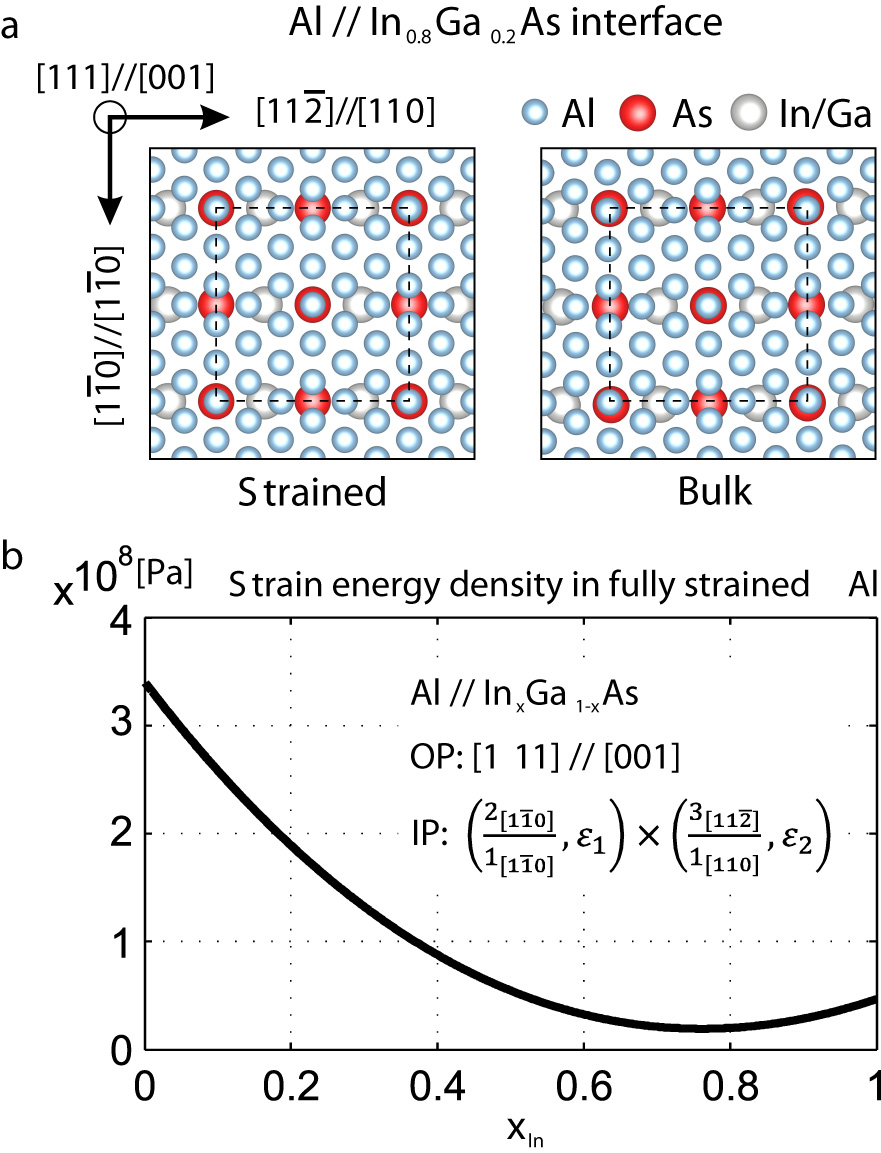}
\caption{(Color online) (a) Normal view on the interfacial domain match of the strained and matched Al-InGaAs interface with x=0.8, and for comparison we show the relaxed and separated phases in bulk. (b) The interface strain energy calculation due to domain matching in the direction of $[1\bar{1}0]_{\rm{InGaAs}}$ shows a minimum near $x \sim 0.8$, which is close to the actual composition of our films.}
\end{figure}

We have calculated the interface strain energy density (with $\epsilon^{2} = \epsilon_{1}^{2}+\epsilon_{2}^2+2\nu\epsilon_{1}\epsilon_{2}$ , where $\epsilon_{1}$ and $\epsilon_{2}$ are the strain in the two in-plane directions), for all domain matching combinations of in-plane lattice planes of up to 20 and as a function of $x$, and found that the strain energy density is lowest for the low order lattice plane ratios, 3/2 and 5/2 interfacial matching along $[1\bar{1}0]_{\rm{InGaAs}}$ and $[11\bar{2}]_{\rm{InGaAs}}$ in-plane directions, respectively, with a minimum achieved near $x \sim 0.8$, as shown in Figure S1(b). Note the corresponding ratio of atoms contributing with broken bonds for single plane interfaces are 2/1 and 3/1 which is used as a notation in the figure. 
Since it is more likely to find lower chemical bonding energy on a larger scale for smaller interfacial domains, supporting our assumption for the small 2/1 and 3/1 interfacial domain matching. This match may play an important role for the formation of an overall ordered film with only two degenerate grain variants.

\section{Derivation of Tunneling Hamiltonian}
In this supplementary material section, we provide the details of the derivation of tunneling Hamiltonian at the interface of our semiconductor (Sm) - superconductor (S) system. 

In order to obtain the conduction band profile corresponding to the heterostructure shown in Fig. 2a, we numerically solve coupled Poisson and Schroedinger equations using effective mass approximation. We input the layout of the heterostructure, band offsets, effective electron masses~\cite{vurgaftman2001} and electron sheet density and calculate self-consistently the potential energy profile and the electron density distribution as a function of $z$ coordinate. This task can be accomplished with any of the several Schroedinger-Poisson solvers available such as nextNano (www.nextnano.com). We find that the potential energy profile in the Sm can be approximated with the asymmetric barriers shown in Fig.~S2 with $E_F-U_1 \approx 0.023$~eV and $E_F-V_L \approx 0.12$~eV. The energy difference between the conduction band of InGaAs and the Fermi level at the point of contact to Al ($z_3$) is set to 0.04~eV~\cite{AdachiBook}. The same value is also used as a boundary condition in the numerical simulation.

\begin{figure}[htp]
\centering
\includegraphics[scale=0.25]{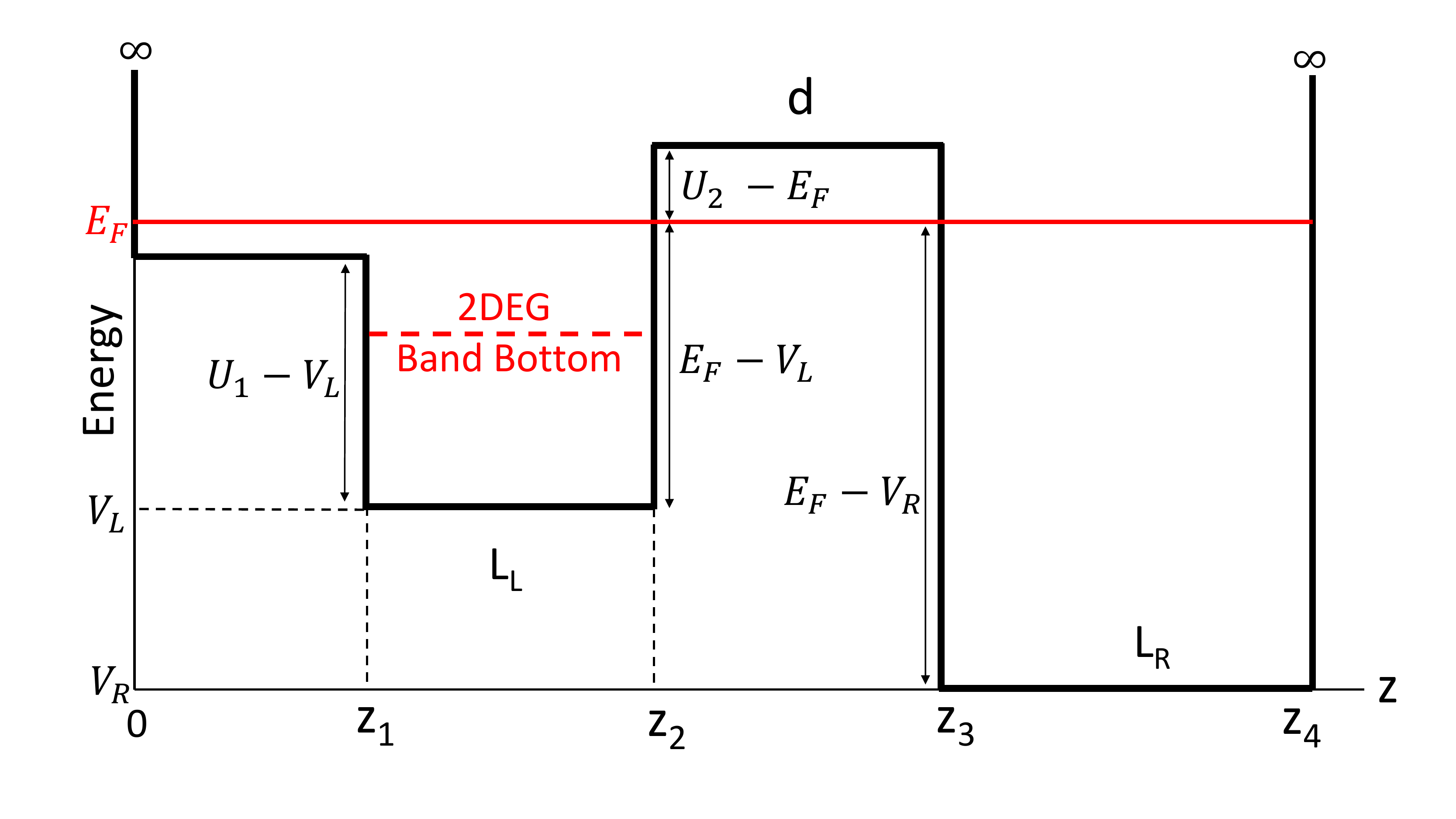}
\caption{ A simplified model of our quantum well structure near the surface}
\label{fig:s2}
\end{figure}

To simplify the analytical calculation, we assume that our system consists of four parts as shown in Fig.~\ref{fig:s2}: InAs quantum well (QW) with potential energy $V_L$ and effective mass $m_L$ for $z_1<z<z_2$, InGaAs barrier with effective mass $m_B$ and potential $U_1$ for $0<z<z_1$, InGaAs barrier with potential $U_2$ for $z_2<z<z_3$, and metallic Al with potential $V_R$ and effective mass $m_R$ for $z_3<z<z_4$. We assume infinite potential energy for $z<0$ and $z>z_4$. We represent the thickness of InAs as $L_L=z_2-z_1$, thickness of Al as $L_R=z_4-z_3$, and the thickness of the middle barrier as $d=z_3-z_2$. We consider QW as a two-dimensional system with a single band (partially occupied) which has a band bottom below $U_1$ and $U_2$ and Al as three dimensional metal. We can describe this system by the many body Hamiltonian:

\begin{equation} 
H=H_{\rm InAs}+H_{\rm Al}+\sum_{\vec{k},\vec{q}}t(\vec{k}, \vec{q})c^\dagger_\vec{k}d_\vec{q} + \text{h.c.}
\end{equation} 

where $c^\dagger_\vec{k}$ is the creation operator of an electron in QW with two dimensional momentum $\vec{k}$ and $d_\vec{q}$ is the annihilation operator for a quasiparticle with momentum $\vec{q}$ in Al. Here tunneling amplitude near the fermi energy can be calculated from [\onlinecite{Bardeen61, PradaSols}],
\begin{equation} 
t(\vec{k},\vec{q})=-\frac{1}{2m_B}\int dxdy \left[ \psi^*_\vec{k}\partial_z \chi_\vec{q} -\chi_\vec{q}\partial_z\psi^*_{\vec{k}} \right]_{z=z_b}
\end{equation} 
where we set $\hbar=1$, $z_b$ is any point inside the $U_2$ barrier, and $\psi_\vec{k}$ and $\chi_\vec{q}$ are the eigenstate wavefunctions of semi-infinite Hamiltonian $H_L$ and $H_R$:
\begin{eqnarray} 
H_L&=&
\begin{cases}
-\frac{\nabla^2}{2m_B}+ U_1, & \,\,0<z<z_1\\
-\frac{\nabla^2}{2m_L}+ V_L, & \,\,z_1<z<z_2 \\
-\frac{\nabla^2}{2m_B}+ U_2, & \,\,z_2<z
\end{cases}\\
H_R&=&
\begin{cases}
-\frac{\nabla^2}{2m_B}+ U_2, & \,\,z<z_3 \\
-\frac{\nabla^2}{2m_R}+ V_R, & \,\,z_3<z<z_4
\end{cases}
\end{eqnarray} 
For two semi-infinite metals separated by the barrier with the height $U$, the tunneling matrix element is given by
\begin{align}\label{eq:tunneling}
t(\vec{k},\vec{q})\approx \frac{4}{\sqrt{v_Lv_R}}\frac{k_zq_z}{\nu_F}\sqrt{\frac{E_F}{U}}e^{-k_Bd}\delta(k_\parallel-q_\parallel)
\end{align}
Here $v_L$ and $v_R$ are the normalization constant proportional to the volume of two sides, $\nu_F=k_F^3/4\pi^2E_F$ is the density of states at the Fermi energy, and $k_B\approx\sqrt{2mU}$ [\onlinecite{PradaSols}]. Eq.\eqref{eq:tunneling} is valid for a system with a uniform effective mass and barrier height much larger than the Fermi energy $U \gg E_F$.

We now derive an appropriate formula for tunneling matrix elements for the system shown in Fig. S2 by taking into account different masses in InAs, InGaAs and Al as well as the finite-size effects associated with the confinement of InAs and Al. Using a similar method for calculating tunneling matrix elements, one finds that the tunneling amplitude in the WKB approximation is given by :
\begin{equation} 
t(\vec{k},\vec{q})=\frac{8\pi^2}{\sqrt{C_L C_R}}\frac{k_{B2}}{m_B}e^{-k_{B2}d}\delta(k_\parallel-q_\parallel).
\end{equation} 
Taking into account the energy and in-plane momentum conservations, one can assume that
\begin{equation} 
k_{B2}=\sqrt{2m_B(U_2-E_F)+k_\parallel^2}=q_B.
\end{equation} 
The normalization constants are given as
\begin{eqnarray} 
C_L&=&A\left(\frac{2k_LL_L+\sin2\phi-\sin2(k_LL_L+\phi)}{2k_L\sin^2(k_LL_L+\phi)}+\frac{1}{k_{B2}}+\frac{(1-4e^{-k_{B1}z_1}+(3+2k_{B1}z_1)e^{-2k_{B1}z_1})\sin^2\phi}{k_{B1}(1-e^{-k_{B1}z_1})^2\sin^2(k_LL_L+\phi)}\right)   \\
C_R&=&A\left(\frac{L_R}{\sin^2(q_RL_R)}+\frac{1}{q_{B}}+\frac{m_R q_{B}}{m_B q_R^2}\right)
\end{eqnarray} 

where $A$ is the area of the interface, $\tan\phi=(1-e^{-k_{B1}z_1})k_Lm_B/k_{B1}m_L$, $k_{B1}=\sqrt{2m_B(U_1-E_F)+k_\parallel^2}$, $k_L$ and $q_R$ are the quantized wavevectors along $z$-direction in InAs and Al which satisfy the following boundary conditions $\tan (k_LL_L+\phi)=-k_Lm_B/k_{B2}m_L$ and $\tan(q_RL_R)=-q_Rm_B/q_Bm_L$ imposed by the continuity equation for $\psi_\vec{k}'(z)/m(z)$ and $\chi_\vec{q}'(z)/m(z)$. 

The escape rate from the two-dimensional (2D) InAs into the bulk metal can be calculated using Golden Fermi rule:
\begin{equation}
\Gamma(\vec{k})=2\pi \sum_{\vec{q}} |t(\vec{k},\vec{q})|^2 \delta(E_L-E_R)
\end{equation}
Assuming that the Fermi wave length in Al is much smaller than the thickness $L_R$, one can replace the summation over $q_R$ with an appropriate integration $\sum_{\vec{q}}=A L_R \int \frac{d^3 q}{(2\pi)^3}$ yielding
\begin{eqnarray} 
\Gamma(\vec{k}) = \int_{-\infty}^{\infty} d q_z \frac{4L_R A^2}{C_L C_R}\frac{k_{B2}^2}{m_B^2}e^{-2k_{B2}d}\delta(E_L-E_R)=\frac{8L_R A^2}{C_L C_R}\,\frac{m_Rk_{B2}^2}{m_B^2q_R}e^{-2k_{B2}d}
\end{eqnarray} 
Since $\Gamma(\vec{k})$ does not depend on the direction of $\vec{k}$ and most of the contribution comes from near the Fermi energy, we can assume $\Gamma(\vec{k})\approx\Gamma(k_F)\equiv \Gamma$.
Having derived the dependence of the escape rate from the Sm to S on microscopic parameters, we now consider that the optimization problem of the superconducting proximity effect. After integrating out superconducting degrees of freedom, the effective low-energy Hamiltonian for the Sm becomes~\cite{Stanescu2011}:
\begin{align}
H_{\rm eff}=(1+\Gamma/\Delta_0)^{-1}H_{\rm InAs}\tau_z+\Gamma(1+\Gamma/\Delta_0)^{-1}\tau_x
\end{align}
where $\tau_a$ are Pauli matrices acting on the Nambu space and $\Delta_0$ is the bulk gap in Al. When $\Gamma/\Delta_0 \gg 1$, the induced superconducting gap is large, of the order of $\Delta_0$ but effective $g$-factor and spin-orbit coupling are suppressed. In the opposite limit, the induced gap $\Delta_{ind}$ is small $\Delta_{ind} \sim \Gamma \ll \Delta_0$. Clearly, the optimal regime corresponds to $\Gamma \sim \Delta_0$ when electrons spend equal amount of time in the Sm and S.

The presented derivation assumes tunnelling regime and should thus only be used for thick enough barriers. The data presented in Fig. 1b was calculated based on the material parameters known in literature~\cite{vurgaftman2001,AdachiBook} and the electron sheet density in the Sm of $0.5\times10^{12}~$cm$^{-2}$. The predictions can be made more quantitative by using the material parameters and electron density measured for a certain device. Alternatively, the two parameters that define the exponential dependence of the tunnelling rate on the barrier thickness for given materials (the prefactor and the exponential factor) can also be found using experimental measurements data for devices with different $d$ but identical otherwise.

\section{Critical field of Al thin films}

In this section we present the Chandrasekhar-Clogston theory~\cite{Chandrasekhar1962, Clogston} description of the temperature dependence of the critical field \cite{Wojcik2015}:

\begin{equation}
\log\left(\frac{\Delta (T = 0)}{\Delta (T)}\right) = \large\int_{1}^{\infty} dx \frac{1}{\sqrt{(x^{2}-1)}} f_{F}\left(x\frac{\Delta(T)}{T}\right)
\end{equation}

where $f_{F}$ is the Fermi distribution function. By solving the above integral numerically and relating $\Delta_{T}$ to $B_{c}$ in a linear form, the data of Fig.~3c in the main text are fitted to Eq. S13 as shown in Fig.~S3.

\begin{figure}[htp]
\centering
\includegraphics[scale=0.375]{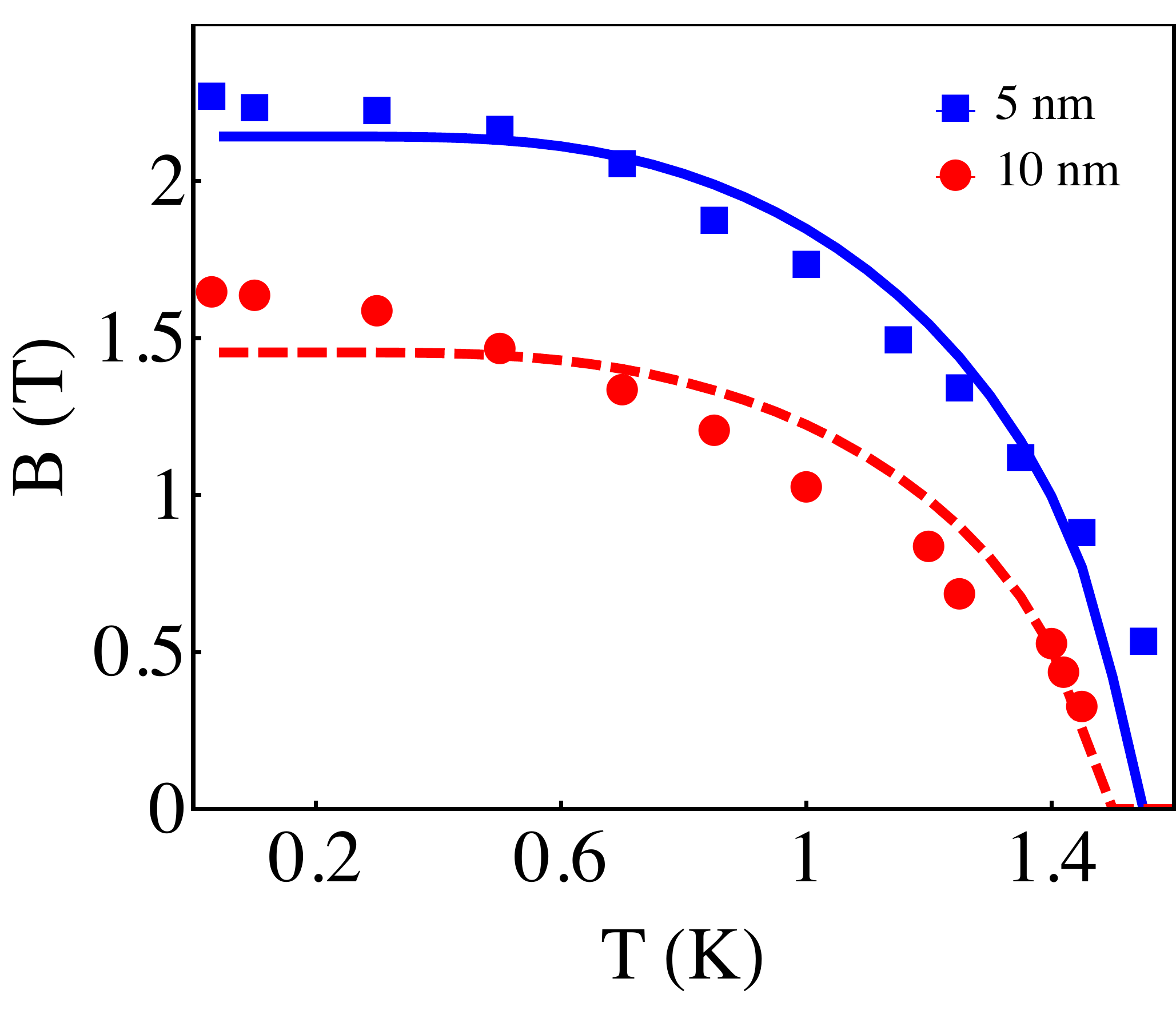}
\caption{(Color online) The Chandrasekhar-Clogston fit to Al thin film data}
\end{figure}

\section{S-Sm-S device fabrication and measurement}

\begin{figure}[htp]
\centering
\includegraphics[scale=0.6]{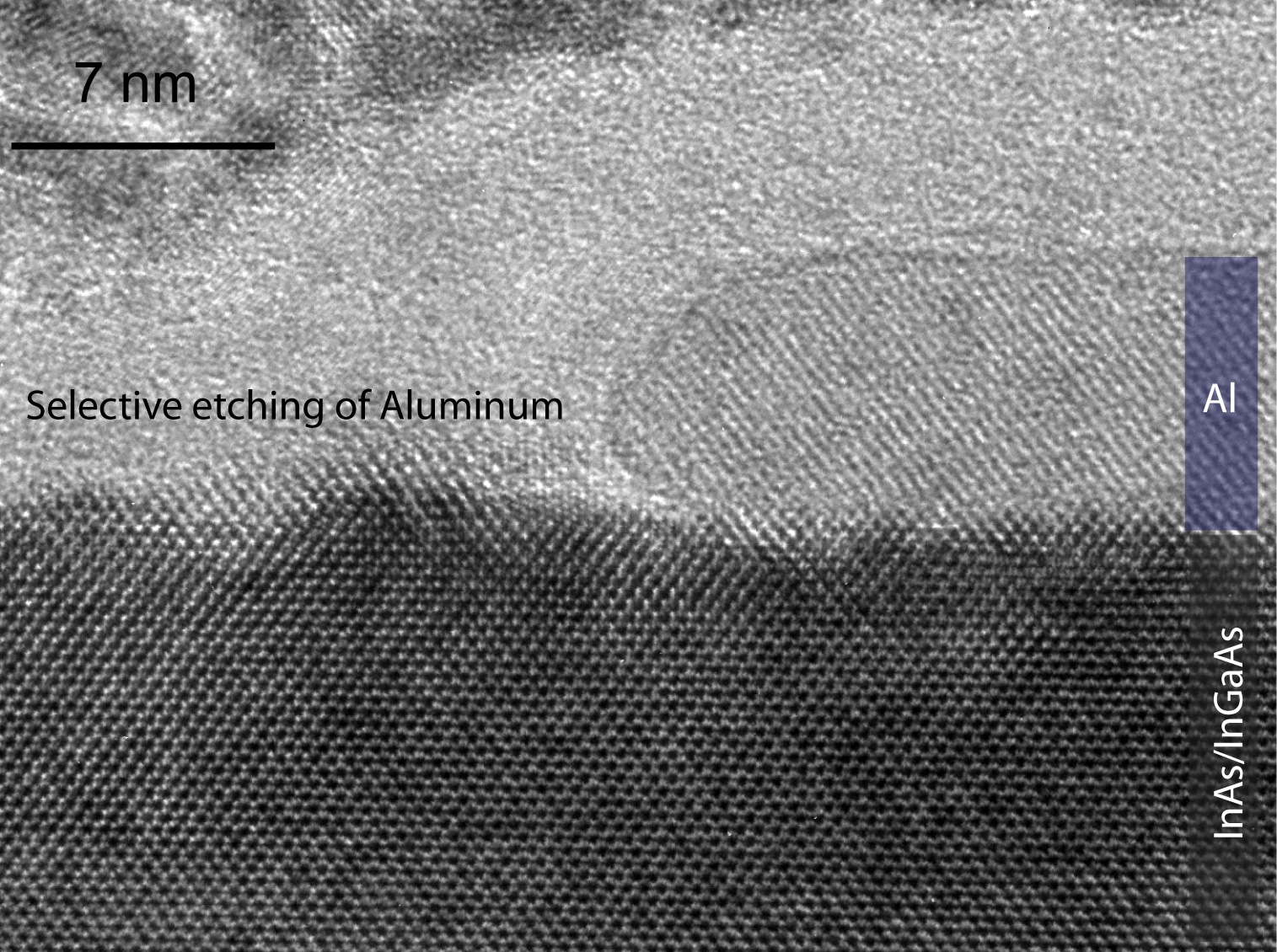}
\caption{(Color online) High resolution TEM image of a processed sample with selective etching of Al over InGaAs.}
\end{figure}

Hybrid S-Sm devices are fabricated using selective etching of Al over InGaAs material. Diluted HF, Transene Al etchant Type D and controlled BCl$_{3}$ dry etch have all shown successful removal of Al. Figure~S4 shows the high resolution TEM image of a processed sample. The atomically sharp and flat interface is preserved and Al removal procedure has not damaged the flat Sm surface.

\begin{figure}[htp]
\centering
\includegraphics[scale=1.05]{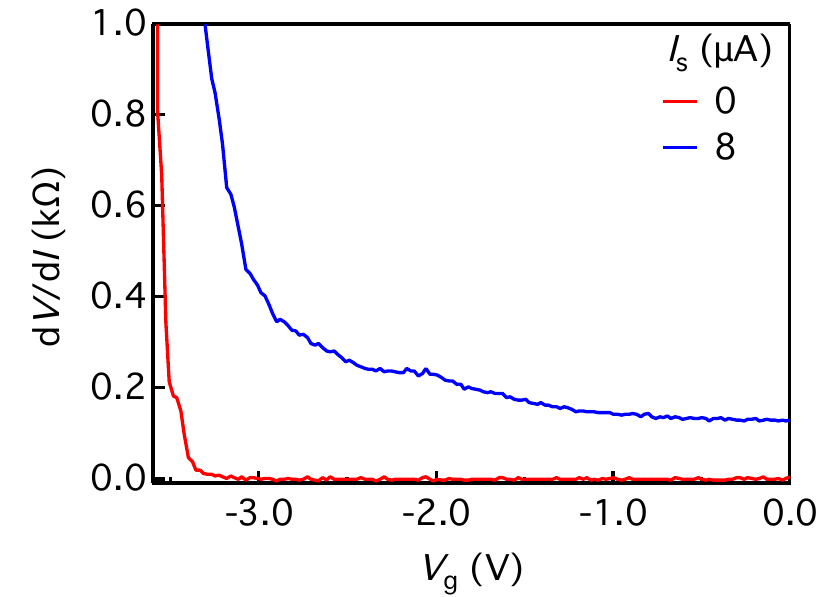}
\caption{(Color online) Resistance of the S-Sm-S junction at zero (red) driving current and 8$~\mu$~A (blue). At $I_{s} = 8~\mu$~A the the junction is driven normal and hence the resistance behavior is dominant by the Sm conduction.}
\end{figure}

Measurements were made in a dilution refrigerator at the base temperature, T = 30~mK, using four-point ac lock-in techniques with 5 nA excitation current for weak antilocalization measurements and sourcing bias voltages ranging from 10$~\mu$V to 100$~\mu$V for conductance measurements. Atomic layer deposition of 40~nm of Al oxide creates the dielectric for the top Ti/Au finger gate.

Figure S5 shows the resistance of normal ($I_{s} = 8~\mu$~A in blue) and superconductivity induced ($I_{s} = 0~\mu$~A in red) as a function of finger gate bias. The sharp rise in normal resistance signals the depletion of the 2DES in the Sm at $V_{g} < -3$~V.  As the conduction in the Sm is decreased, the induced superconductivity in the junction disappears evidenced by disappearance of the critical current and rise of zero-resistance state.


\end{document}